%% file: main.tex
\documentclass[sigconf]{acmart}

\AtBeginDocument{%
  \providecommand\BibTeX{{%
    \normalfont B\kern-0.5em{\scshape i\kern-0.25em b}\kern-0.8em\TeX}}}

\setcopyright{acmcopyright}
\copyrightyear{2018}
\acmYear{2018}
\acmDOI{10.1145/1122445.1122456}

\acmConference[Woodstock '18]{Woodstock '18: ACM Symposium on Neural
  Gaze Detection}{June 03--05, 2018}{Woodstock, NY}
\acmBooktitle{Woodstock '18: ACM Symposium on Neural Gaze Detection,
  June 03--05, 2018, Woodstock, NY}
\acmPrice{15.00}
\acmISBN{978-1-4503-XXXX-X/18/06}

\usepackage{graphicx}
\usepackage{amsmath}
\usepackage{xcolor}
\usepackage{soul}
\usepackage{enumitem}
\usepackage{pifont}% http://ctan.org/pkg/pifont
\usepackage{multirow}
\usepackage{booktabs}
\usepackage{mathtools}

\usepackage{enumitem}
\usepackage{multirow}
\usepackage{booktabs}
\usepackage{mathtools}
\usepackage{placeins}[section]

\usepackage{balance} % for balancing columns on the final page

\newcommand{\ournet}{EEF1-NN}
\newcommand{\floor}[1]{\lfloor #1 \rfloor}

\title[]{EEF1-NN: Efficient and EF1 allocations through Neural Networks}

\author{Shaily Mishra}
\affiliation{%
  \institution{International Institute of Information Technology (IIIT)}
  \city{Hyderabad}
  \country{India}}
\email{shaily.mishra@research.iiit.ac.in}

\author{Manisha Padala}
\affiliation{%
  \institution{International Institute of Information Technology (IIIT)}
  \city{Hyderabad}
  \country{India}}
\email{manisha.padala@research.iiit.ac.in}

\author{Sujit Gujar}
\affiliation{%
  \institution{International Institute of Information Technology (IIIT)}
  \city{Hyderabad}
  \country{India}}
\email{sujit.gujar@iiit.ac.in}

\begin{abstract}
Neural networks have shown state-of-the-art performance in designing auctions, where the network learns the optimal allocations and payment rule to ensure desirable properties. Motivated by the same, we focus on learning fair division of resources, with no payments involved. Our goal is to allocate the items, goods and/or chores efficiently among the fair allocations. By fair, we mean an allocation that is Envy-free (EF). However, such an allocation may not always exist for indivisible resources. Therefore, we consider the relaxed notion, Envy-freeness up to one item (EF1) that is guaranteed to exist. However, it is not enough to guarantee EF1 since the allocation of empty bundles is also EF1. Hence, we add the further constraint of efficiency, maximum utilitarian social welfare (USW). 
In general finding, USW allocations among EF1 is an NP-Hard problem even when valuations are additive.  In this work, we design a network for this task which we refer to as \ournet.\  We propose an  UNet inspired architecture, Lagrangian loss function, and training procedure to obtain desired results. We show that \ournet\  finds allocation close to optimal USW allocation and ensures EF1 with a high probability for different distributions over input valuations. Compared to existing approaches \ournet\ empirically guarantees higher USW. Moreover, \ournet\ is scalable and determines the allocations much faster than solving it as a constrained optimization problem. 
\end{abstract}

\keywords{Fairness, Efficiency, and Neural Networks}

\newcommand{\BibTeX}{\rm B\kern-.05em{\sc i\kern-.025em b}\kern-.08em\TeX}

\begin{document}

%%% The following commands remove the headers in your paper. For final 
%%% papers, these will be inserted during the pagination process.

\pagestyle{fancy}
\fancyhead{}

%%% The next command prints the information defined in the preamble.

\maketitle 

%%%%%%%%%%%%%%%%%%%%%%%%%%%%%%%%%%%%%%%%%%%%%%%%%%%%%%%%%%%%%%%%%%%%%%%%

\section{Introduction}
\label{sec:intro}

\input{introduction}

%%%%%%%%%%%%%%%%%%%%%%%%%%%%%%%%%%%%%%%%%%%%%%%%%%%%%%%%%%%%%%%%%%%%%%%%

\section{Related Work.}
\label{sec:relatedwork}
Fair resource allocation is well studied in the literature across various fairness and efficient notions \cite{endriss2018lecture,bouveret_chevaleyre_maudet_moulin_2016, MarkakisTRENDS2017,procaccia_moulin_2016,inbook}. When a definition of fairness is too strong or may not exist, we always look for its relaxation/approximation; researchers have also studied how likely it is that a fairness notion will not exist. 

In this paper, we are majorly concerned with EF1 and USW. 
EF1 allocations always exist and can be found in polynomial time. When agents have additive valuations, the round-robin algorithm always guarantees EF1 for (pure) goods or chores and the double round-robin algorithm for the combination of goods and chores. \cite{Caragiannis2018,Caragiannis2016}. When agents have general valuations, we can find EF1 allocations in $O(mn^3)$ using a cycle-elimination algorithm. \cite{Lipton2004}. Finding MUW allocations is also polynomial-time solvable for additive valuations, i.e., we iterate over items, assign the item to the agent who values it the most.  However, finding MUW allocation amongst EF1 allocations is NP-hard even for two agents with additive valuations \cite{aziz2019constrained,Barman2019Nphard,ef1inumHaris,Aziz2016AAMAS,Caragiannis2018,KeijzerBKZ09}. 

 Authors in \cite{Highmultiplicitypaper2019} present a framework to compute $\epsilon$-Efficient and $\mathcal{F}$-Fair allocation, using parametric integer linear programming, which is double exponential in terms of the number of agents and items. In \cite{Highmultiplicitypaper2019}, they explore group Pareto Efficiency, which is equivalent to USW. Authors in  \cite{ef1inumHaris} provide a pseudopolynomial-time algorithm for finding MUW within EF1 for any fixed number of agents for goods, which is exponential in the number of agents. In the paper, \cite{Iwillhaveorder}, the authors present an approximately optimal round-robin order that gives highly efficient (USW) EF1 allocations in the Reviewer Assignment setting; however, the setting is quite different from ours, as we are not concerned with the multiplicity of items.

In further related work, papers \cite{barman2018finding,Caragiannis2018} explore, PE and EF1 allocations, \cite{EQ1freemangoods,EQ1ChoresFreeman} explore PE and EQ1 allocations, \cite{azizPRop1andfpo} explore PE and Prop1 allocations for various items (goods or/and chores). There will always be a tradeoff between fairness and efficiency, corresponding to the study of the price of fairness \cite{barmanpriceoffairness,priceoffairnessbeiijcai19}.
Alongside, Researchers have also studied how likely it is that a fairness notion will not exist \cite{closinggaps,dickersonThecomputationalriseandfalloffairness,whendoenvyfreeallocationexist}. In \cite{closinggaps}, the authors show that Round Robin allocation is envy-free when $m \ge \Omega(n \log n / \log \log n)$.

Recently the EconCS community has been interested in learning mechanisms/algorithms using neural networks, esp. in a setting of theoretical limitations.
For, e.g., In paper \cite{Noah2018}, the authors provide a strategy-proof, multi-facility mechanism that minimizes expected social cost via NN. Authors in \cite{ICAByML}, the authors integrate machine learning in the combinatorial auction for preference elicitation. Further, in \cite{ICAByDL}, authors use a neural network to improve it and reformulate WDP into a mixed-integer program. Authors in \cite{manisha2018learning,TacchettiDeepmind} learn optimal redistribution mechanisms through NNs. Another line of work is Reinforcement Mechanism Design, such as learning dynamic price in sponsored search auctions ~\cite{Reinforcement2,Reinforcement1}.  In \cite{PublicprojectNN}, the authors use NN to maximize the expected number of consumers and the expected social welfare for public projects.

% \sm{TBD-add citation,and finish nn related work}
%%%%%%%%%%%%%%%%%%%%%%%%%%%%%%%%%%%%%%%%%%%%%%%%%%%%%%%%%
\section{Preliminaries}
\label{sec:prelim}
We consider the problem of allocating $M=[m]$ indivisible items among
$N=[n]$ interested agents. Each agent $i \in N$ has a valuation function $v_i : 2^{M} \rightarrow \mathbb{R}$ and $v_i(S)$ is its valuation for a $S \subseteq M$ s.t. $v_i(\phi) = 0$. We consider three settings - pure goods, pure chores, and a combination of goods and chores. In combination, an item may be good for one agent and a chore for another. For an agent $i$, an item $j \in M$ is a \emph{good} if, $v_i(\{j\})\geq 0$, and a \emph{chore} if, $v_i(\{j\}) < 0$. We represent valuation profile $v=(v_1,v_2,\ldots,v_n)$. We consider additive valuations. The valuation of an agent $i \in N$ for bundle $A_i$ is $v_i(A_i) = \sum_{j\in A_i} v_i(\{j\}) $. Utilitarian Social Welfare (USW) is defined as $sw(A,v) = \sum_{i\in N} v_i(A_i)$.
We assume $\mathcal{F} = F_1 \times F_2, \ldots, \times F_n$ to be a known prior distribution over agents' valuations. We randomly draw $v_i \sim F_i$. 
An allocation $A \in \{0, 1\}^{n\times m}$ is an $n$ way partition  $(A_1, \ldots A_n)$ of $M$. Here, $A_i \in [m]$ is the bundle assigned to the agent $i$ and $A_i \cap A_k = \phi, \forall i, k \in N$ and $i \neq k$.
We consider a complete allocation of items, i.e., $\cup_i A_i = M$. We use the notation $n \times m$, for a problem setting with $n$ agents and $m$ items.
Given a valuation profile of agents $v=(v_1,v_2,\ldots,v_n)$, our goal is to find a fair and efficient allocation. First, we define fairness and efficiency properties. 

\begin{definition}[Envy-free (EF)]
\label{def:ef}
An allocation $A$ is said to be EF, if no agent envies another agent, i.e., $\forall i,j \in N , v_i(A_i) \ge v_i(A_j)$. 
\end{definition}

As EF allocation may not always exist for indivisible goods, we consider a generalized version of relaxation of the EF defined by Budish~\cite{Budish11}.

\begin{definition}[Envy-free up to one item (EF1)]
\label{def:ef1}
An allocation $A$ is said to be EF1 if envy of any agent can be eliminated by either removing any good from the envied agent’s allocation or removing any chore from the agent’s allocation. i.e., when either of the following is true
$\forall i,k \in N$,
\begin{enumerate}
    \item $ \exists j \in A_k \ \mbox{s.t.}\   v_i(A_i) \ge v_i(A_k \setminus  \{j\})$
    \item $\exists j \in A_i \ \mbox{s.t.}\  v_i(A_i \setminus \{ j\}) \ge v_i(A_k)$
\end{enumerate}
\end{definition}

\begin{definition}[Maximum Utilitarian Welfare (MUW)]
An allocation $A^*$ is said to be \emph{efficient} or MUW if it maximizes the USW, i.e.\\ $$A^* \in \underset{A \in \{0,1 \}^{n\times  m}}{arg\,max} sw(A,v) $$
\end{definition}

\begin{definition}[EEF1 Allocation]
We say an allocation is \emph{EEF1} if it satisfies EF1 fairness and maximizes USW amongst EF1 allocations. 
\end{definition}

%%%%%%%%%%%%%%%%%%%%%%%%%%%%%%

%%%%%%%%%%%%%%%%%%%%%%%%%%%%%%%%%%%%%%%%%%%%%%%%%%%%%%%%%%%%%%%%%%%%%%%%
\section{Our Approach: \ournet}
\balance
\label{sec:eef1-nn}
%%%%%%%%%%%%%%%%%%%%%%%%%%%%%%%%%%%%%%%%%%%%%%%%%%%%%%%
\ournet\ represents a mapping from valuation space to allocation space, i.e., $\mathcal{A}^w :\mathbb{R}^{\{n \times m\}} \rightarrow \{0,1\}^{n\times m}$, where $w$ represents the network's weights. To learn the network parameters, we formulate our problem to optimize social welfare w.r.t. to fairness constraints in Section~\ref{ssec:loss_fn}. We construct the Langrangian Loss function of this optimization problem for the training of \ournet.
We explain our architecture in Section ~\ref{ssec:networkdetails} and training details in Section~\ref{ssec:training}. Note that we represent \ournet\ by $\mathcal{A}^w$ and an allocation by $A$.
\smallskip

% \noindent\textbf{Optimization Problem. }
\subsection{Optimization Problem.}
\label{ssec:loss_fn}
Consider $n$ interested agents and $m$ indivisible items; it can be good $v_i(\{j\}) \ge 0$, or chore $v_i(\{j\}) < 0$. We are given a set of valuation profile $v=(v_1,v_2,\ldots,v_n)$, where $v_i$ is drawn randomly from a distribution $\mathcal{F}_i$.
Among all possible allocations $A \in \{0,1\}^{n \times m}$, we need to find an optimal $A^*$ that maximizes utilitarian social welfare $sw(A, v)$ and satisfies a fairness constraint. 
We formulate two fairness constraints - EF and EF1.
In Definition \ref{def:ef}, the envy of an agent $i \in N$ according to EF is as follows,
\begin{align}
\label{eq:EFenvy}
    envy_i(A,v) =& \sum_{k \in N}\max 
\Bigg\{ 0 ,  (v_i(A_k) - v_i(A_i))  \Bigg\}
\end{align}
In Definition. ~\ref{def:ef1}, the $ef1_i$ of an agent $i \in N$, according to EF1 is as follows:
\begin{align}
\label{eq:E1envy}
    ef1_i(A,v) =& \sum_{k \in N}\max 
\Bigg\{ 0 ,  (v_i(A_k) - v_i(A_i))
+ \nonumber \\ &\min \left\{ -\max_{j \in A_k} v_i(\{j\}) , \min_{j \in A_i} v_i(\{j\}) \right\}  \Bigg\}
\end{align}
The above constraints are generalized formulations for both goods and chores. Given a set of valuation profiles, our goal is to maximize the expected welfare w.r.t. to the expected fairness.

\smallskip
\begin{center}
\fbox{\begin{minipage}{0.8\columnwidth}
\begin{equation}
\label{eq:our_prb}
\begin{aligned}
\mbox{minimize}   & \; \;  - \mathbb{E}_v\left[sw(A,v)\right] =  \mathbb{E}_v[\sum_{i\in N} v_i(A_i)]
\\
\mbox{subject to}  & \; \;  \mathbb{E}_v\left[\sum_{i \in N} envy_i(A,v)\right] = 0 \quad \mbox{or,} \\
                &\; \; \mathbb{E}_v\left[\sum_{i \in N} ef1_i(A,v)\right] = 0
 \end{aligned}
\end{equation}\end{minipage}}
\end{center}
\smallskip

In the above optimization problem, we have 'OR' among fairness constraints, which we will elaborate on in the Ablation Study in Section ~\ref{subsec:ablaS}.

\smallskip
\noindent\textbf{\ournet: Lagrangian Loss Function.}
We now formulate the objective function given by Eq. \ref{eq:our_prb} using the Lagrangian multiplier method. We use the Lagrangian multiplier $\lambda \in \mathbb{R}_{\ge 0}$ to combine the objective and constraints. Given $\mathcal{L}$ samples of valuation profiles $(v^1, \ldots, v^{\mathcal{L}})$ drawn from $\mathcal{F}$, we have the corresponding input $I_v^l$ and the loss for each sample is given by, 
\begin{equation}
    Loss(I_v^l, w, \lambda) =  \frac{1}{n\times m}
\Bigg[ -sw(\mathcal{A}^w(I_v^l),v^l) + 
\lambda\frac{ \sum_{i \in N}  envy_i(\mathcal{A}^w(I_v^l),v^l)}{n}  \Bigg] 
\label{eq:loss_per_sample}
\end{equation}
We minimize the following loss w.r.t $w$,
\begin{equation}
\label{eq:loss}
\mathbf{L}_{EEF1}(I_v^l, w, \lambda)= \frac{1}{\mathcal{L}}\sum_l Loss(v^l,w)
\end{equation}

%%%%%% For worst case if we include modify the above to also include worst case formulation %%%%%%%%%%%%%%%%%%%%%%%%%%%%%%%
%%%%%%%%%%%%%%%%%%%%%%%%%%%%%%%%%%%%%%%%%%%%%%%%%%%%%%%%%%%%%%%%%%%%%%%%
\subsection{Network Details} 
\label{ssec:networkdetails}
We describe \ournet\'s various components, including the input, architecture, and other training details in this section. \ournet\ is a fully convolutional network (FCN) and processes input of varied sizes (i.e., height $\times$ width). Because of using an FCN, \ournet\ runs independently of $n$ and $m$.

\smallskip
\noindent\textbf{\ournet: Input.} 
We construct an input tensor of size $n \times m \times 6$, i.e., the input to the network is a six channeled input $I_v \in \mathbb{R}^{n\times m \times 6}$. The first channel is an $n\times m$ matrix of given valuations, i.e., $\forall i,j ; I_v[i, j, 1] = v_i(\{j\})$. Note that we sample the valuation from a distribution $\mathcal{F}$. 
We take a matrix $X$ of size $n \times m$  that contains valuation for items only corresponding to the agent who values it the most, and the rest elements are zeros, i.e., 
It takes a value $v_i(\{j\})$ for each item (column) for the agent (row) having maximum value for it, i.e, 
% We formulate $X$ as follows, $\forall j \in M$
% $\forall j in M \; ; X[i,j]= {v_i(\{j\})}_{i \in argmax_i v_i(\{j\})} $, otherwise $X[i,j] = 0$
\begin{align*}
\forall j \in M; \; X[i.j,1] &=
  \begin{cases}
   v_i(\{j\})        & \text{if } i \in argmax_i v_i(\{j\}) \\
  0        & \text{otherwise}
  \end{cases}
\end{align*}

We break ties arbitrarily. We expand this $n \times m$ matrix into five channels, such that the first one will contain information about items indexed as $0,5,10,\ldots, \floor{m/5}$, i.e.,
\begin{align*}
 I[i.j,2] &=
  \begin{cases}
   X[i,j,1]        & \text{if } j \in \{0,5,10,\ldots,\floor{m/5}\} \\
  0        & \text{otherwise}
  \end{cases}
\end{align*}
% $I_v[i,j,2] =  v_i(\{j\})_{i \in argmax_i v_i(\{j\}) \cap j \in \{0,5,10,\ldots, m/5\}} $

The next channel will have data from the previous channel and along with that about items indexed as $1,6,11,\ldots, 1+\floor{m/5}$. 
\begin{align*}
 I[i.j,3] &=  I[i,j,2]  + 
  \begin{cases}
    X[i,j,1]  & \text{if } j \in \{1,6,11,\ldots,1+\floor{m/5}\} \\
   0  & \text{otherwise}
  \end{cases}
\end{align*}
And so on. The last channel, $I_v[i,j,6]$ will be equal to $X$
We observe that giving single channeled input of only valuations, i.e., tensor of size $(n \times m \times 1)$, performs sub-optimal as opposed to the six-channel. We evaluate the performance of six channeled input across various other inputs in Section ~\ref{subsec:ablaS}. 
With this representation, the network learns better.
% can converge to an allocation that is closer to the optimal, i.e., fair and efficient. 

\smallskip
\noindent\textbf{\ournet: Architecture.}
Our architecture is inspired by U-Net architecture \cite{unetpaper}. U-Net is a fully convolutional network built to segment bio-medial images; it also requires assigning labels to image patches and not just classifying the image as a whole. Traditionally, a fully convolutional network is used for image segmentation. While we are working on valuation profiles rather than images, one of the primary motivations to use U-Net is to process arbitrary size images. If we use a feed-forward fully functional neural network to learn fair and efficient allocations, we need a different network for each $n \times m$. Moreover, just using a feed-forward functional network (multi-layer perceptron) learns EEF1 allocations for smaller values of $n$, cannot learn as $n$ increases; we will briefly mention this in our Ablation Study in Section ~\ref{subsec:ablaS}.

\begin{figure}[h]
    \centering
     \includegraphics[width=\linewidth]{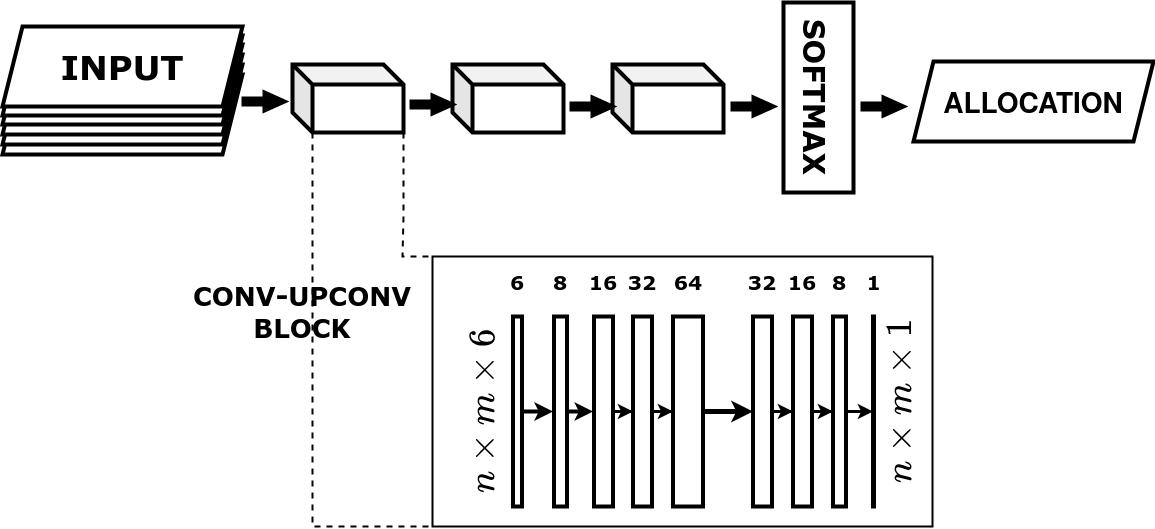}
    \caption{\centering  \ournet\ Architecture}
    \label{fig:EEF1-NNArchitecture}
    \Description{For 10 agents, valuation drawn randomly from a uniform distribution, plotting-TBD }
\end{figure}

\ournet\ contains series of convolution (contracting) and up-convolution (expanding) layers, as given by Fig. \ref{fig:EEF1-NNArchitecture}. \ournet\ has three series of Conv-UpConv layers. 
The convolutional layers consist of 4 repeated 3x3 convolution, each followed by a non-linear activation function, i.e., tanh, which is applied element-wise. 
%\sm{change network once I have experiment running}
The up-convolution layers consist of 4 repeated 3x3 up-convolution, each followed by a tanh activation.
Note that we are not using maxpool or skip connections. We found that using a pooling layer or skip connections degraded the network performance. The final output represents the probability with which agent $i$ will receive item $j$. We apply softmax activation function across all agents for every item to ensure each item is allocated exactly once ,i.e., $ \forall j \in M \;  \sum_{i \in N} \mathcal{A}^w_{i}(\{j\})=1  $. In total, we have 24 layers (convolution + up-convolution). 

Using an FCN structure, we have a generalized network for $n \times m$; however, learning EEF1 allocations is not easy. We need to learn discrete variables, while neural networks are known for learning continuous output. We will describe these challenges in the next Section.
% Once trained with sufficient samples, \ournet\ can be 
% \ournet\ can learn 
% Though we observed that a fully connected linear network learns EEF1 allocations for smaller values of $n$, as $n$ increases, it does not generalize with the increased complexity of finding EEF1. Moreover, we need to train it across all required ($n$, $m$) pairs separately.  \ournet~ is fully convolutional layers; hence it is independent of the input parameters. Once trained with sufficient samples can be tested for any $n$ and $m$. 

% \begin{figure}[!t]
%     \centering
%     \subfloat[\centering  EEF1 Architecture]{{\includegraphics[width=6.5cm]{AAMAS22-NN/figures/EEF1NetArchitecture (1).png} }}%
%     \ \ 
%     \subfloat[\centering Ablation Study over hyper-parameters]{{\includegraphics[width=5cm]{AAMAS22-NN/figures/ablation_study.png} }}%
%     \caption{Network Architecture and Ablation Study}%
%     \label{fig:Netplusablation}
% \end{figure}

%\sm{if doing worst case}%
% We use two step gradient descent to optimize the loss such that $\underset{w}{min}\  \underset{\lambda}{max}\  \mathbf{L}_{EEF1}(I_v^l, w, \lambda) $.

%%%%%%%%%%%%%%%%%%%%%%%%%%%%%%%%%%%%%%%%%%%%%%%%%%%%%%%%%%%%%%%%%%%%%%%%
\subsection{Training Details}
\label{ssec:training}
% To train we use Adam Optimizer \cite{Adamopt} with fixed learning rate $0.001$. 
There are certain challenges with network training, especially in the setting of indivisible resource allocation. 
% The network 
% We use a temperature factor along with our softmax, as we want discrete allocation, i.e. for indivisible items. 

\noindent\textbf{Integral Allocations. }  The global optima of the optimization problem in Eq. \ref{eq:our_prb} might lie in a continuous allocation setting, i.e., similar to allocating divisible items. The training starts if a network learns to distribute an item equally among all agents, and the gradient vanishes. For a fair allocation, assigning an equal partition of each item is indeed an optimum. Converting these non-integral probabilities to integral allocation is non-trivial. Hence we set a \emph{temperature} parameter in the softmax layer of the network to prevent getting stuck at such optima.
 Let $p_{j} = \{p_{j_1}, \ldots, p_{j_n} \}$ denote the output of our network before the final layer which represents the probability of assigning item $j$ to all the agents. The final allocation for agent $i$, i.e., $\mathcal{A}^w_i(\{j\})$is given by,
$$ \mathcal{A}^w_i(\{j\}) = \mbox{softmax}(p_{j_i}) =  \frac{e^{p_{j_i}/T}}{\sum_{k=1}^n e^{p_{j_k}/T}}$$

It is common to start with a large temperature value for initial exploration and gradually reduce the temperature to reach the global optima. 
While training, when we set the temperature value to 1, we get fractional allocations.
As we decrease the value of $T$, the network outputs allocation close to discrete. The approach we want is while training, allocation output is almost discrete, but not exactly discrete. When we keep the value of $T$ too low, the output is exact discrete allocations, and there is no learning because of the vanishing gradients  \cite{elibendersky_2016}. So we appropriately choose $T$ based on our experiments. Once the network learns, we set the parameter low enough to ensure discrete allocations. 
% We believe our network architecture of having series of Conv-UpConv layers also contributes  towards this.

% We are trying to escape global optima
%     % i.e. assigning allocation in continuous settings,
%     and interested in a set of local optima. Let $p_{j} = \{p_{j_1}, \ldots, p_{j_n} \}$ denote the output of our network before the final layer which represents the probability of assigning item $j$ to all the agents. The final value for agent $i$ i.e., $\mathcal{A}^w_i(\{j\})$is given by,
% $$ \mathcal{A}^w_i(\{j\}) = \mbox{softmax}(p_{j_i}) =  \frac{e^{p_{j_i}/T}}{\sum_{k=1}^n e^{p_{j_k}/T}}$$
% It is common to start with a large value of temperature for initial exploration and gradually reduce the temperature over time to reach the global optima. In our setting, as we decrease the value of $T$, the network outputs allocation close to discrete and vice-versa. 
% The approach we want is while training output is almost discrete, but not exactly discrete. When we keep the value of $T$ too low, the output is exact discrete allocations, and there is no learning because of the vanishing gradients \cite{elibendersky_2016}. When the value of $T$ is too high, the allocation is not close to integral, and it converges near the global optima. So we appropriately chose $T$ for a setting based on our experiments. Once the network learns, we set the parameter low enough to ensure discrete allocations. 

\noindent\textbf{Inefficient Local Optima. } Due to the low-temperature value, the training of \ournet\ is highly unstable and often gets stuck at inefficient local optima. To overcome this, we use the technique of \emph{Bootstrap Aggregation} or Bagging \cite{bagging}. It combines the predictions from multiple classifiers to produce a single classifier. Hence we train multiple weak networks with varied hyper-parameters on the same data set, capturing different sets of local optima. While testing, the final allocation is aggregated from these networks. We pass a test sample through all networks and select the allocation that is EF1 with maximum USW. 
%We observe that this instability in network training is based on Lagrangian multiplier $\lambda$. And we overcome this unstable training via aggregating the network results. 
Usually, in Bagging, we train neural networks with different training data sets, however in our case, varying $\lambda$ produces different models. In total, we bag seven networks with varied $\lambda \in [0.1,2]$ for increased performance. We further analyze how Bagging affects our results in the ablation study provided in the experimental section.  
%\sm{TBDHERE - how bagging}

% \subsubsection{Training Details} 
We implement \ournet\ using Pytorch. 
The network weights are initialized using Xavier Initialization \cite{xavierinit}. To train, we use Adam Optimizer \cite{Adamopt} with fixed learning rate $0.001$. We use a batch size of 256 samples. We sample valuations from  $U[0,1]$ (goods), $U[-1, 0]$ (chores) and $U[-1,1]$ (combination). We sample $150k$ training data for both $10 \times 20$ and $13 \times 26$ for goods, chores, and combinations, so in total, we have $300k$ training samples, and we sample $10k$ testing samples for each setting. We transform these valuations into six-channeled input and feed to the network. We set the temperature parameter to 0.01. We train our network for 1000 epochs. We use our Lagrangian loss as the objective function to train our network. We train seven networks with varied $\lambda \in [0.1,2]$ and bag them for enhanced performance. The training process took 5-6 hours to train a single network using GPU.
We are training the network for $10 \times 20$ and $13 \times 26$. however, we show our test results for various $n \times m$. We test for network performance for $n \in [7,15]$. Further, we also train a individual network over different distributions such as Gaussian ($\mu$=0.5,$\sigma$=1), Log-normal ($\mu=0.5$,$\sigma=1$), and Exponential ($\lambda = 1$), with $150k$ training samples for $n=10$.

% Note that we cannot run this network for $n <10$, as we are applying series of convolution and up-convolution. Hence we reduce a layer of convolution and up-convolution from our Conv-UpConv block as shown in Fig. ~\ref{fig:EEF1-NNArchitecture}. We trained this reduced network with valuations equally samples from $7 \times 14, 10 \times 20,$ and $13 \times 26$. We observed that reducing this layer affects performance for $n \ge 10$, so we use this network to test only for $n<10$.

% \sm{TBDHERE - n=7,10,13}

% \sm{we train our network for 0.001 for 100 epochs and then 0.0001 for rest of the epochs}

%%%%%%%%%%%%%%%%%%%%%%%%%%%%%%%%%%%%%%%%%%%%%%%%%%%%%%%%%%%%%%%%%%%%%%%%
\section{Experiments and Results}
\label{sec:experiemtnsandresults}
For reporting the network performance on the test set, we define the following two metrics, one is the measure of fairness and the other of efficiency.

\smallskip
\noindent\textbf{Evaluation Metrics. }
\begin{enumerate}[noitemsep, leftmargin=*]
\item $\alpha_{EF1}^{ALG}$ - It measures the probability with which an algorithm $ALG$ outputs, EF1 allocation. $\alpha_{EF1}$ is the ratio of the number of samples that are EF1 to the total number of samples.
\item $\beta_{SW}^{ALG}$ - It measures the ratio of expected USW of an algorithm $ALG$ by expected USW of MUW allocation. $\beta_{SW}^{ALG} =  \mathbb{E}(sw^{ALG}) / \mathbb{E}(sw^{MUW})$.
\end{enumerate}

Using the above metrics, we conduct the following ablation study to set appropriate hyper-parameters.
% Our network is trained with $300k$ valuations sampled from a pre-specified distribution, i.e., for uniform distribution, we train the network with $300k$ samples (goods, chores, combination) of $n=10$ and $300k$ of $n=13$.
Our network is trained across various types of items (goods or/and chores) and types of distribution.  
The test set consists of $10k$ samples. 
Note that $\beta^{ALG}_{SW} \in [0,1]$ for goods, $\beta^{ALG}_{SW} \ge 1$ for chores, and will depend on the overall social welfare (positive/negative) for a combination of goods and chores. 
We will use that notation $(\alpha_{EF1},\beta_{sw})$ to write network/algorithm's performance.

%%%%%%%%%%%%%%%%%%%%%%%%%%%%%%%%%%%%%%%%%%%%%%%%%%%%%%%%%%%%%%%%%%%%%%%%
\subsection{Ablation Study}
\label{subsec:ablaS}
\begin{figure}[h]
    \centering
    \includegraphics[width=\linewidth]{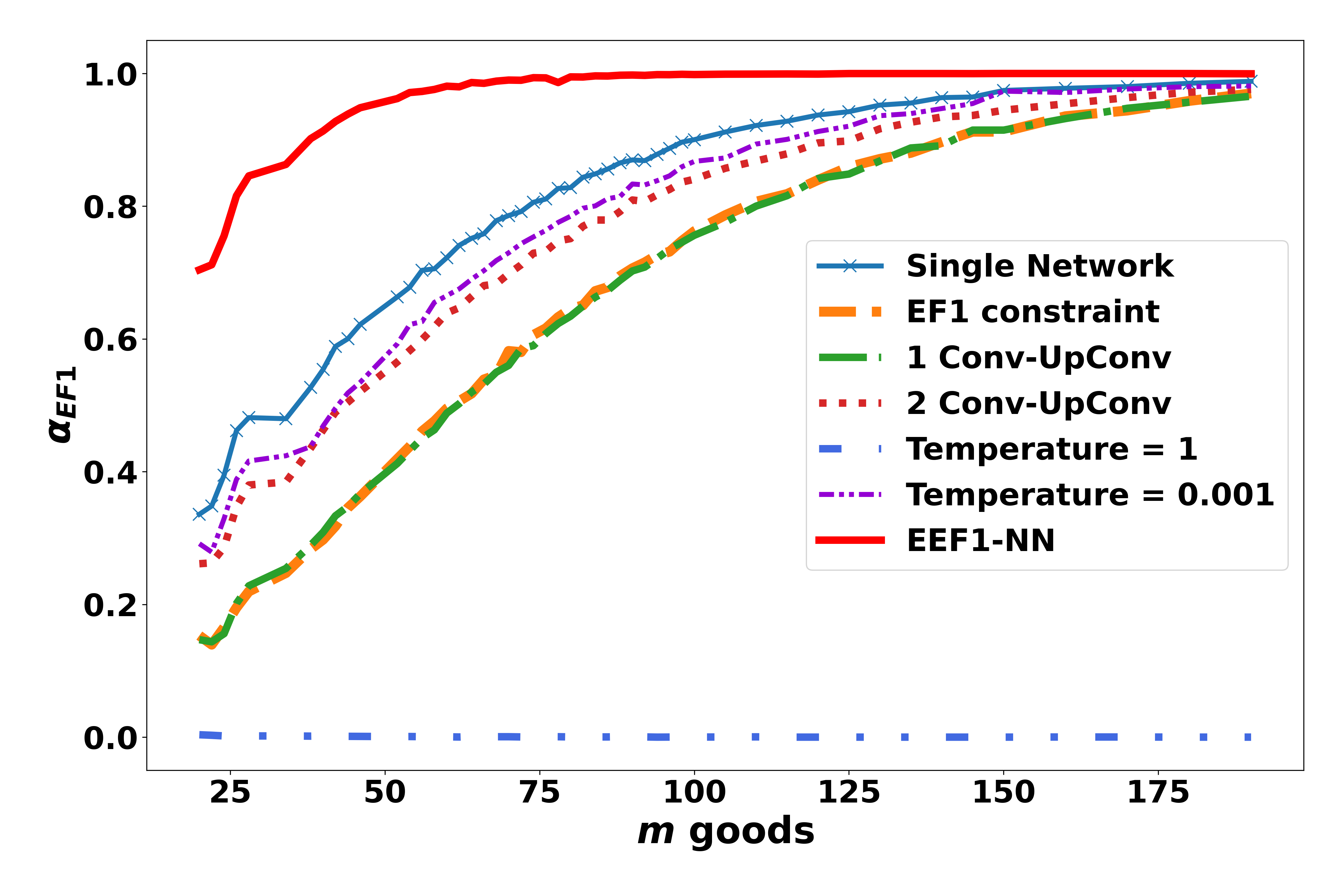} 
    \caption{Ablation Study over hyper-parameters}
    \label{fig:ablationstudy}
    \Description{For 10 agents, valuation drawn randomly from a uniform distribution, plotting-TBD }
\end{figure}

\begin{figure}[h]
    \centering
    \includegraphics[width=\linewidth]{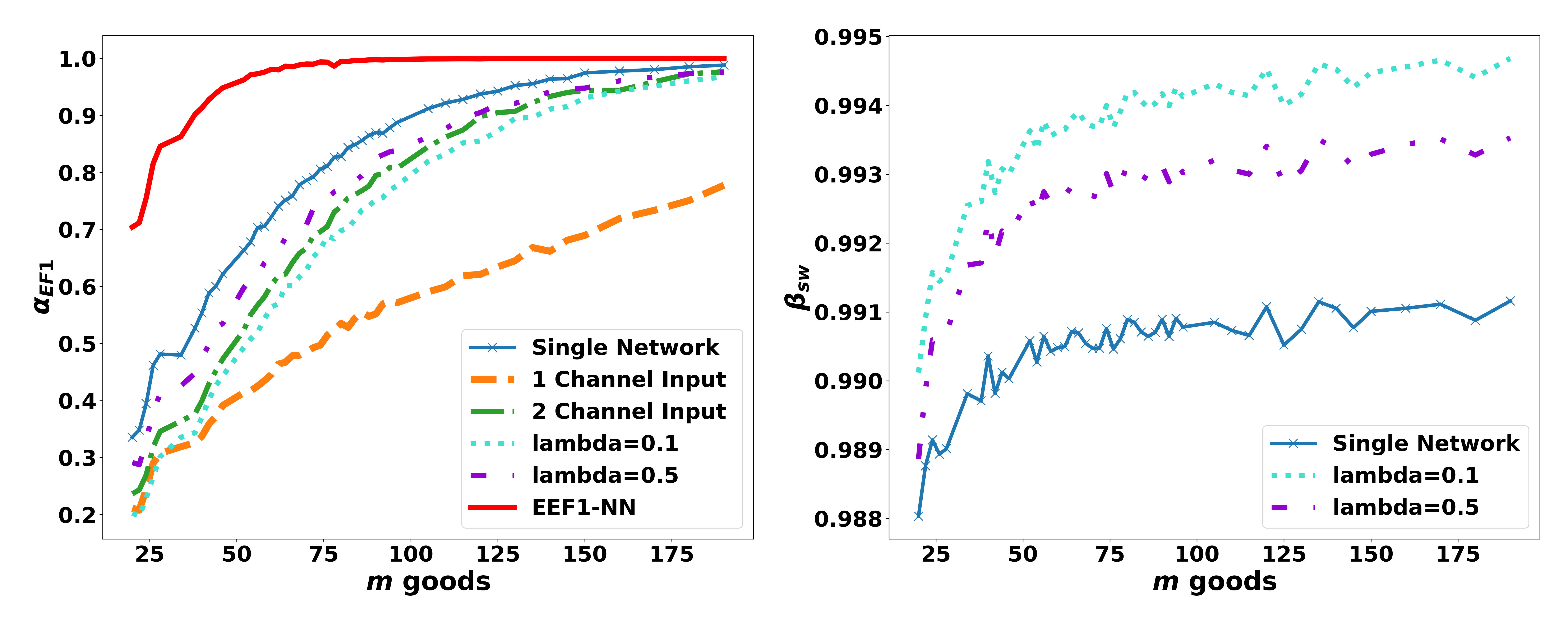} 
    \caption{Ablation Study over Input channels and $\lambda$}
    \label{fig:ablationstudy2}
    \Description{For 10 agents, valuation drawn randomly from a uniform distribution, plotting-TBD }
\end{figure}

We illustrate the effect of specific hyper-parameters in the performance of \ournet\ in Fig. [\ref{fig:ablationstudy},\ref{fig:ablationstudy2}]. 
We sample the valuations from the uniform distribution, set $n=10$, only goods, for all the ablation study experiments, and observe the $\alpha_{EF1}$ as $m$ increases. 
In the plots, the red line with the label \ournet\ denotes the $\alpha_{EF1}$ for optimal parameters.
Corresponding to \ournet, a single network from this bagged network is labeled as \emph{Single Network} in the graph. This \emph{Single Network} trained with six-channeled input, $\lambda=1$, and temperature $T=0.01$ is the baseline to compare across this ablation study. Only one parameter is changed w.r.t. the \emph{Single Network} training for all the networks used in this study. 
% For all the ablation study experiments, we sample the valuations from uniform distribution and set $n=10$, and observe the $\alpha_{EF1}$ as $m$ increases. 
We only compare $\beta_{sw}$ for the network across varied $\lambda$ values, as $\alpha_{EF1}$ values are close to the \emph{Single Network}.

\smallskip
\noindent \emph{(i) Effect of Temperature $T$. } 
  Keeping other parameters fixed, we vary the $T=\{1,0.1,0.001\}$ in Fig. \ref{fig:ablationstudy}. When $T=1$, our network converges to global optima, i.e., fractional allocation, unable to learn EEF1 discrete allocation represented by the blue line at the bottom of the plot. Also, we empirically observed that when networks learn to allocate an equal fraction of an item among agents (0.1 in case of 10 agents), the gradient vanishes, thus stuck in a bad local optimum. When $T=0.001$ (violet line), it is too low, and performs sub-optimally compared to when $T = 0.01$ in single network (no bagging) (light blue line). We also noticed that the network's performance for $T=0.01$ and $T=0.1$ are close to each other. We set $T=0.01$ for all the bagged networks in \ournet.
  
\smallskip
\noindent\emph{(ii) Effect of series of Conv-UpConv layers.} We select three series of Conv-UpConv For \ournet\  as illustrated in Fig. \ref{fig:ablationstudy}. We plot the $\alpha_{EF1}$ for one Conv-UpConv series green dashed line. It is less than what we obtain for a 2-series red dotted line, which is less than the optimal 3-series (light blue line).  As seen from Fig. \ref{fig:ablationstudy}, an increase between 1-series and 2-series is significant compared to 2-series and 3-series (single network without bagging). The complexity of the network having 4-series is far more than the performance improvement. We have limitations over the number of layers in Conv-UpConv, as we are working with a low dimensional matrix, such as $10\times 20$, and having such a series increases performance. Briefly stating, while training a 4-layered feed-forward fully functional network for $10 \times 20$, $\alpha_{EF1}$ was roughly 0.008.

% \noindent \emph{(i) Effect of Temperature $T$. } 
 
\smallskip
\noindent \emph{(iii) Effect of loss function }
We analyze how different envy definitions in our loss function represented in Eq. \ref{eq:our_prb} affects the training of \ournet.  As shown in Fig. \ref{fig:ablationstudy}, when we train our network using EF, i.e., Eq. \ref{eq:EFenvy} (\emph{Single Network}, light blue line), the network performs significantly better than when trained using EF1, i.e., Eq.\ref{eq:E1envy} (orange dashed line).
For example, for $10 \times 20$, the performance of \emph{Single Network} is (0.3358,17.9611), whereas the performance of the EF1 trained network is (0.1530,17.8708). Given a distribution, one way of interpreting this behavior can be that when we train the neural network to maximize social welfare w.r.t. to Envy-free, the best fairness it can have is EF1 while maintaining efficiency close to MUW. 
% For more information, we can refer to the price of fairness literature. \cite{} 

\smallskip
\noindent \emph{iv) Number of Input Channels. }
When training with just one channel input, i.e., valuations, the results we obtained were quite poor; for $10\times20$, we get $(0.2113,17.8976)$ as shown in Fig~\ref{fig:ablationstudy2}. Thus we changed our input representation into more channels. We tested for 2,6, and $(n+1)$ channels for $n=10$. For two channeled input, we set the first channel of input tensor as the valuation matrix and the second channel to matrix $X$, described in Section ~\ref{ssec:networkdetails}. Like six-channeled input, we expand the matrix $X$ to 11 channels, i.e., the number of channels here is equal to $n+1$. The learning of a six-channeled network, i.e., \emph{Single Network} is better than the two-channeled network. The performance of a two-channeled network is $(0.2365,17.8991)$, \emph{Single Network} is $(0.3358,17.9611)$, and 11-channeled network is $(0.3925,17.9395)$. We didn't plot $\alpha_{EF1}$ of the 11-channeled network because Even though $\alpha_{EF1}$ is high of the 11-channeled network, the network is dependent on $n$; cannot be generalized for $n$, along with an increase in input representation complexity.

\smallskip
\noindent \emph{v) Effect of Bagging. }
We try different combinations of networks, each trained for varied $\lambda$ values. The Lagrangian Loss, as described in Eq. ~\ref{eq:loss_per_sample}, $\lambda$ corresponds to the fairness constraint. More the value of $\lambda$, more penalty is given to envy in the loss. When $\lambda$ is too small, the penalty for allocating an unfair allocation is less, so the network learns a more efficient but less fair allocation. As we increase $\lambda$ up to a certain value, the network learns less efficient but more fair allocations. Beyond a certain value, if we increase $\lambda$, we get a degraded performance overall. In Fig. ~\ref{fig:ablationstudy2}, we compare $\alpha_{EF1}$ and $\beta_{sw}$ of \emph{Singe Network} ($\lambda$=1), lambda=0.5, and lambda=0.1.
We observed that varying $\lambda$ value results in converging to the different optimum. 
We bagged seven networks trained on with $\lambda$ values of $\lambda \in [0.1,2]$. We choose a mix of (low efficiency, high fairness) and (high efficiency, low fairness). We feed six-channeled input to \ournet, and it outputs the fairest and efficient, i.e., if more than one network gives EF1 allocation, then it will select the one with maximum social welfare.
In Fig. \ref{fig:ablationstudy}, we find combining the networks (red line) outperforms the performance of a single network (light blue line).
% \sm{TBDHERE}

% For this instance, we select the model consisting of four networks with $\lambda=1$, two networks with $\lambda=5$, and one network with $\lambda=10$.

% We conducted experiments with several settings to evaluate \ournet\
\subsection{Experiment Details and Observations}
We select the best training parameters for \ournet\ based on the above ablation study for the following experiments.  We conduct three types of experiments to compare existing approaches across, \textsc{Exp1}: Different kinds of resources, \textsc{Exp2:} Different input distributions, and \textsc{Exp3:} Scalability to samples with large $n$. In all three experiments, we compare \ournet\ with the following existing methods,
% We designed three experimental set-ups to validate \ournet \\ \textbf{\textsc{Exp1}}:Performance for Uniform Distribution for goods, chores, and combination with for $n=10$\\%: In this we study $\alpha_{EF1}^{ALG}$ and $\beta_{sw}^{ALG}$ for $n=10$.  We study the effect of $m$ on these performance metrics when the valuations are additive and generated from uniform distribution. \\
% \textbf{\textsc{Exp2}}: Performance for $n=10$ for various distribution.\\ 
% \textbf{\textsc{Exp3}} : Performance for $n=13$ for goods, and chores, and for $n=14$ for goods.

% Figures [\ref{fig:n10},\ref{fig:dist},\ref{fig:n13}] are our observations for \textsc{Exp1}, \textsc{Exp2}, and \textsc{Exp3} respectively. 

% \subsubsection{Experimental Results and Discussions}
\begin{itemize}[noitemsep, leftmargin=*]
\item \noindent \textsc{MUW}
Since we don't have Optimal EEF1 allocations to compare our results, we compare our results with MUW allocations. We also analyze after which value of $m$, an MUW allocation is likely to be EF1.
    \item \noindent 
\textsc{Round Robin (RR)} \cite{Caragiannis2016} finds EF1 allocations under additive valuations  for pure goods and pure chores. \emph{Double Round Robin }\textsc{(D-RR)} \cite{Caragiannis2018} extends RR to find EF1 allocation for the case with a combination of goods and chores under additive valuations.

\item \noindent \textsc{CRR}
Based on paper \cite{aziz2019constrained}, we implement CRR to find RB sequence such that it allocate items to the agent who values it the most for goods. As mentioned in \cite{aziz2019constrained}, an RB sequence is EF1 when all items have positive valuations.

\end{itemize}
Note that approaches like using parametric ILP solver \cite{Highmultiplicitypaper2019} and the algorithm provided by \cite{ef1inumHaris} are exponential. Therefore, it is infeasible to use these for the configurations we report our results on, so we do not compare them. 

Further, in Table ~\ref{tab:mvalueanalysis}, we study the convergence of different approaches towards EEF1 for uniform distribution, i.e., after which value of $m$, a method converges to EEF1.
\smallskip
% \subsubsection{}{\textbf{}

\begin{figure*}[!tbh]
  \centering
  \includegraphics[width=0.9\linewidth]{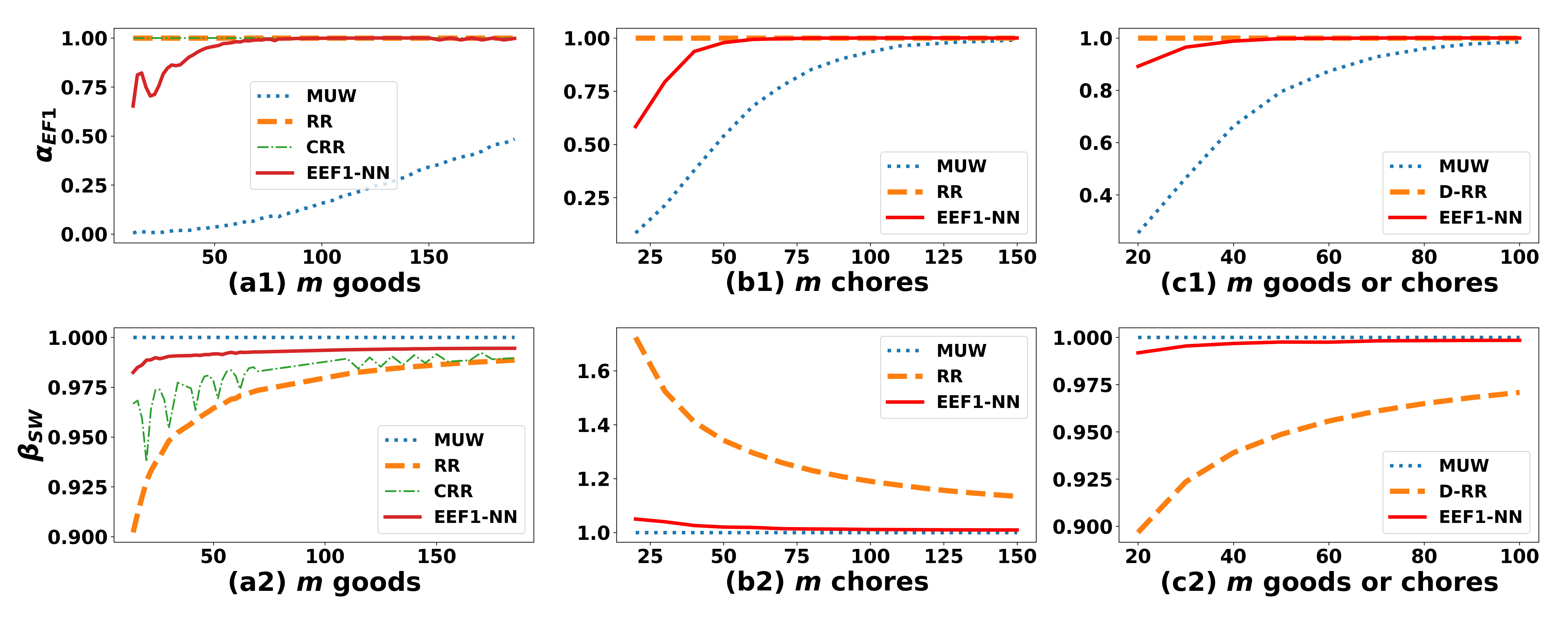}
  \caption{\textsc{Exp1} ($n=10$, Uniform Distribution)}
\label{fig:n10}
\Description{For 10 agents, valuation drawn randomly from a uniform distribution, plotting }
\end{figure*}

\noindent\textbf{\textsc{EXP1:} Performance across differed resources for Uniform Distribution.} For $n=10$, we compare the $\alpha_{EF1}$ in Fig. \ref{fig:n10} (a1, b1, c1) and $\beta_{SW}$ in Fig. \ref{fig:n10} (a2, b2, c2) as $m$ increases across the existing approach. 

Irrespective of the resource type, as the number of items increases, all the approaches will move closer to EEF1. We observe that MUW allocation (blue dotted line) converges towards EEF1 allocations much faster for chores or combinations than goods.
While Round Robin converges to EEF1 allocations much faster in goods compared to chores or combinations. We discuss this convergence in detail in Table ~\ref{tab:mvalueanalysis}. 
We observe that \ournet\ consistently has better $\alpha_{EF1}$ than MUW allocation and better $\beta_{sw}$ than RR/CRR.

The baseline \textsc{RR} is designed to find EF1 allocations and hence, by construct has $\alpha_{EF1}^{RR} = 1$. We observe that $\alpha_{EF1}^{EEF1-NN}$ is close to $\alpha_{EF1}^{RR}$ after a certain $m$. At the same time, the allocation returned by \ournet\ is far more efficient than $RR$ as represented by the $\beta_{SW}$ values. (Fig \ref{fig:n10} (a2,b2,c2)). Though $\alpha_{EF}^{CRR} = 1$ (Fig. \ref{fig:n10} a1), note that the baseline \textsc{CRR} only works for goods.  Even for goods, we observe that compared to \textsc{CRR}, \ournet\ obtains marginally better $\beta_{SW}$, in Fig. \ref{fig:n10}(a2).

% In Fig. \ref{fig:n10}, we plot $\alpha_{EF1}$ and $\beta_{SW}$ for different $m$. We find that MUW allocation is likely to be like EF1 for $m \ge 500$ for goods, i.e., $\alpha_{EF1}^{MUW}$ is 0.96, $m \ge 150$ for chores, i.e. $\alpha_{EF1}^{MUW}$ is 0.99 , and $m \ge 80$, i.e. $\alpha_{EF1}^{MUW}$ is 0.99  for a combination. Since WCRR,RR, and D-RR give EF1 allocations, in Fig. \ref{fig:n10} $\alpha_{EF1}$ is 1. We observe that $\alpha_{EF1}^{\ournet}$ is more than 0.98 for $m \ge 55$, and more than 0.99 for $m \ge 70$ in Fig. \ref{fig:n10}(a1). In Fig. \ref{fig:n10}(a2) $\beta_{sw}^{\ournet}$ is almost is more than 0.99 for $m \ge 30$. While $\beta_{sw}^{WCRR}$ is more than 0.99 for $m \ge 160$, and $\beta_{sw}^{RR}$ is more than 0.99 for $m \ge 200$. We found that for goods, MUW converges to EF1 for a larger $m$, in comparison to RR converging towards MUW, i.e. for much smaller $m$ than 500, RR $\beta_{sw}^{RR}$ is almost 0.99 of MUW. For chores, in Fig. \ref{fig:n10}(b1), $\alpha_{EF1}^{\ournet}$ is more than 0.98 for $m \ge 50$, and more than 0.99 for $m \ge 70$. $\beta_{sw}^{\ournet}$ is almost 1.01 for $m \ge 50$, while $\beta_{sw}^{RR}$ is almost 1.01 for $m \ge 200$ in Fig. \ref{fig:n10}(b2). We observed that RR converges towards MUW after much larger $m$, in comparison to MUW converging to EF1. For combination, in Fig. \ref{fig:n10}(c1), $\alpha_{EF1}^{\ournet}$ is almost 0.99 $m \ge 40$. In Fig. \ref{fig:n10}(c2), $\beta_{EF1}^{\ournet}$ is almost 0.99 for $m \ge 20$, while $beta_{EF1}^{D-RR}$ is 0.98 for $m = 150$. 

% \begin{figure}[h]
%     \centering
%     \includegraphics[width=\textwidth]{AAMAS22-NN//figures/10_agents.png}
%     \caption{\textsc{Exp1} ($n=10$, Uniform Distribution)}
%     \label{fig:n10}
% \end{figure}

\begin{figure*}[!htb]
    \centering
    \includegraphics[width=0.9\linewidth]{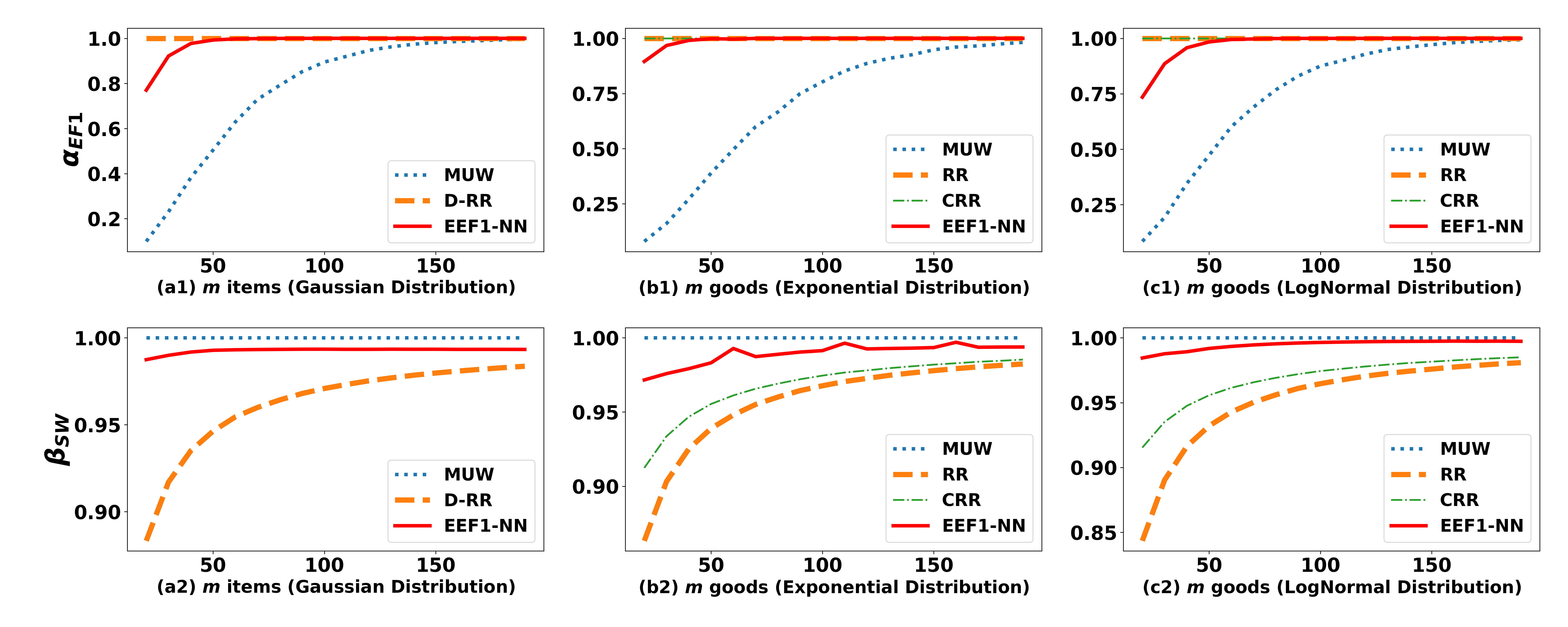}
    \caption{\textsc{Exp2} ($n=10$, different distributions)}
    \label{fig:dist}
\Description{For 10 agents, valuation drawn randomly from a uniform distribution, plotting-TBD }
\end{figure*}

\smallskip
\noindent\textbf{\textsc{EXP2:} Performance across different distributions.}
We provide the performance of \ournet\  when the valuations are sampled from different distributions such as Gaussian ($\mu$=0.5,$\sigma$=1) in Fig \ref{fig:dist}(a1, a2), Log-normal ($\mu=0.5$,$\sigma=1$) in Fig \ref{fig:dist}(b1, b2), and Exponential ($\lambda = 1$) in Fig \ref{fig:dist}(c1, c2). Note that when we sample valuations from Gaussian distribution, it corresponds to allocating a combination of goods and chores. From Fig. \ref{fig:dist}, we observe that in all three distributions, $\alpha_{EF1}^{\ournet}$ is more than 0.99 and $\beta_{SW}^{\ournet}$ is more than 0.99 for $m \ge 40$.

\begin{figure*}[!htb]
    \centering
    \includegraphics[width=0.9\linewidth]{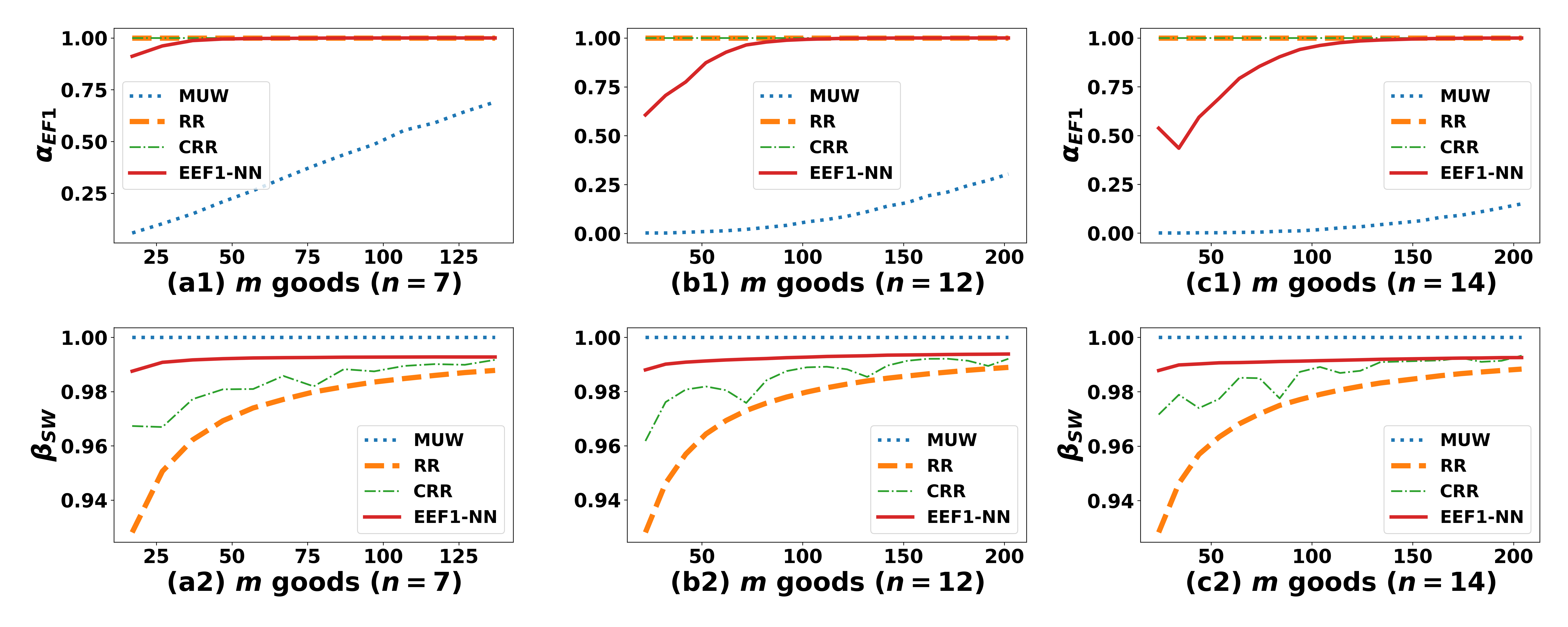}
    \caption{\textsc{Exp3} ($n=7,12,14$ goods, Uniform Distribution)}
    \Description{For 10 agents, valuation drawn randomly from a uniform distribution, plotting-TBD }
    \label{fig:n13}
\end{figure*}

\smallskip
\noindent\textbf{\textsc{EXP3:} Scalability to larger number of agents.} 
\ournet\ is trained only for $10\times20$ and $13\times26$.
As we have seen in the previous results and in Fig. \ref{fig:n13}, the performance scales across varying $m$ seamlessly. In this section, we provide the performance of \ournet\ when $n=7$, $n=12$, and $n=14$ in Fig. \ref{fig:n13}. 
Our network will not run when $n < 10$, given the four $3\times3$ Convolution-UpConvolution. Hence to report performance for $n \in [7,9]$, we reduce a Convolution-UpConvolution layer from our network and train accordingly with $7\times14$ and $10\times20$ valuation profiles.
\ournet\ scales appropriately across $n$; however, if we test the network performance of $n=14$ and $n=20$, the network performs better for $n=14$, solely because we trained just using $10\times20$ and $13\times26$. For a much higher value of $n$, we need to expand our training samples.  

\smallskip
\noindent\textbf{Analysis of Convergence to EEF1 Allocations (Uniform Distribution)}
\begin{definition}[$m^{\star}(n)$]
For a given  $n$, we say an algorithm converges to EEF1 allocation at $m^{\star}(n)$ if $\forall m >m^{\star}(n)$, \\(i) \emph{For goods}: $\alpha_{EF1}^{ALG} \ge 0.99$ and $\beta_{sw}^{ALG} \ge 0.99$. \\(ii) \emph{For chores}: $\alpha_{EF1}^{ALG} \ge 0.99$, and $\beta_{sw}^{ALG} \le 1.02$.
\end{definition}
We empirically study the value of $m^{\star}(n)$ after which \ournet, RR, and MUW start converging towards EEF1 for goods/chores for uniform distribution in Table ~\ref{tab:mvalueanalysis}. We don't report CRR in this Table; as we see fluctuations in $\beta_{sw}$, it doesn't increase smoothly in Fig. [\ref{fig:n10}\ref{fig:n13}];  However, note that CRR results better than RR for goods, and \ournet\ performs marginally better than CRR.  We observe that in the case of goods, \ournet\ reaches close to EEF1 allocations faster than MUW and RR, and RR reaches close to EEF1 faster than MUW.
% (again by convergence in goods, we mean $\alpha_{EF1}^{ALG} \ge 0.99$ and $\beta_{sw}^{ALG} \ge 0.99$).   

% In the case of chores, as seen in Table ~\ref{tab:mvalueanalysis}, \ournet\ converges first, then MUW, and lastly RR. 
As seen in Table ~\ref{tab:mvalueanalysis}, EEF1-NN converges first, then MUW, and finally RR in the case of chores.
We report the value of $m^{\star}$ for RR we use $\beta_{sw} \le 1.064$ since these $m$ values are already significantly high than MUW and RR, concluding that RR will converge after a considerably larger $m$.
We also observed as $m$ increases, $\alpha_{EF1},\alpha_{EFX}$, and $\alpha_{EF}$ of MUW keeps getting closer. For example, for $9\times530$ goods uniform distribution, $\alpha_{EF1}=0.989$,$\alpha_{EFX}=0.9834$, and $\alpha_{EF}=0.9834$; while for $9\times200$ goods uniform distribution, $\alpha_{EF1}=0.6436$,$\alpha_{EFX}=0.5086$, and $\alpha_{EF}=0.5032$.

Note that the actual value of $m^{\star}(n)$ may be slightly different from the exact point of convergence mentioned in Table as we do not perform experiments for all possible values of $m$. Our goal here is to observe a pattern among approaches to compre the different approaches to achieve EEF1.

\begin{table}[H]
\caption{Value of $m^{\star}(n)$ as different approaches converge to EEF1 allocations }
\centering{
\begin{tabular}{l|lll|lll}
\toprule
%   $\lambda$    & $\beta_{SW}$  & $\alpha_{EF1}$  \\
\multicolumn{1}{c}{\multirow{2}{*}{$n$}} & \multicolumn{3}{c}{ ($m$) Goods} & \multicolumn{3}{c}{($m$) Chores}  \\
\multicolumn{1}{c}{}                        &           \ournet\ & RR  & MUW  & \ournet\     & RR  & MUW        \\
\midrule
7                                                & 38                          & 159                    & 380                     & 44                          &  195 &112      \\
8                                                & 46                          & 172                    & 450                     & 44                          & 240& 120       \\
9                                                & 57                          & 186                    & 530                     & 53                          &295  &130     \\
10                                               & 70                          & 196                    & 610                     & 60                          & 340 &148       \\
11                                               & 82                          & 206                    & 660                     & 68                          &400 & 160       \\
12                                               & 94                          & 214                    & 740                     & 75                          & 455 &167       \\
13                                               & 110                         & 220                    & 840                     & 83                          &505  &180       \\
14                                          & 134      &  228   &   940 & 87     & 565         &  190  \\
\bottomrule
\end{tabular}}
\label{tab:mvalueanalysis}
\end{table}

\smallskip
\noindent\textbf{Discussion}
$\alpha_{EF1}^{\ournet}$ reaches 1 much faster than $\alpha_{EF1}^{MUW}$, and $\beta_{sw}^{\ournet}$ reaches close to $\beta_{sw}^{MUW}$ much faster than RR, D-RR, CRR. So the results from \ournet\ show a better trade-off between EF1 and efficiency than the existing approaches for different input distributions. 
% We also observe that even with training only on specific $n$ and $m$ along with goods or/and chores, the performance scales for any $m$ and a large $n$.  
We trained our network with fixed $n \times m$ and goods or/and chores, still the performance scales for any $m$ and a large $n$.
For smaller $n$ and $m$, one can use integer programming or any pseudo-polynomial approach, and when $m>>n$. We observed that MUW converges towards EEF1 faster than RR in goods, while in chores, it's the other way around.
% MUW is EF1 with very high probability, which \ournet\ also mimics it. 
Hence we conclude that our approach effectively learns and provides a better trade-off when $m$ is not too large or very small compared to the $n$ but is in a specific range.

% \sg{mention for small values of n,m one can may use ILP or some existing alogirhtms..but very large value of m given n, one may use MUW (e.g. if n=10, m>200) but in beween, NN are effective in learning algorithm to get EEF1 allocations}

% \begin{figure}%
%     \centering
%     \subfloat[\centering n=13 goods $\alpha_{EF1}$]{{\includegraphics[width=5cm]{ICONIP21/figures/n13goodsaEf1.png} }}%
%     \qquad
%     \subfloat[\centering n=13 good $\beta_{sw}$]{{\includegraphics[width=5cm]{ICONIP21/figures/n13goodsswrationwithmuw.png} }}%
%     \caption{2 Figures side by side}%
%     \label{fig:example}%
% \end{figure}

% \begin{figure}%
%     \centering
%     \subfloat[\centering n=13 chore $\alpha_{EF1}$]{{\includegraphics[width=5cm]{ICONIP21/figures/n13choresaEF1.png} }}%
%     \qquad
%     \subfloat[\centering n=13 chore $\beta_{sw}$]{{\includegraphics[width=5cm]{ICONIP21/figures/n13swratio.png} }}%
%     \caption{2 Figures side by side}%
%     \label{fig:example}%
% \end{figure}

% \begin{figure}%
%     \centering
%     \subfloat[\centering n=13 comb $\alpha_{EF1}$]{{\includegraphics[width=5cm]{ICONIP21/figures/n=13,comb,aEF1.png} }}%
%     \qquad
%     \subfloat[\centering n=13 comb $\beta_{sw}$]{{\includegraphics[width=5cm]{ICONIP21/figures/n=13,comb,b.png} }}%
%     \caption{2 Figures side by side}%
%     \label{fig:example}%
% \end{figure}

%%%%%%%%%%%%%%%%%%%%%%%%%%%%%%%%%%%%%%%%%%%%%%%%%%%%%%%%%%%%%%%%%%%%%%%%
\section{Conclusion}
In this paper, we addressed finding fair and efficient allocations for goods, chores, or combinations. In general, the problem is NP-hard. We proposed a neural network  \ournet\  to find EEF1 allocations.  To overcome the issue of finding optimal discrete allocations, we designed appropriate architecture and input representation combined with other training heuristics. We studied the effect each proposed constituent has on performance. Our experiments demonstrated the efficacy of \ournet\ for different input distributions over existing approaches. It finds reasonably fair and close to optimal solutions in real-time. Can we improve it further?

%%%%%%%%%%%%%%%%%%%%%%%%%%%%%%%%%%%%%%%%%%%%%%%%%%%%%%%%%%%%%%%%%%%%%%%%

%%%%%%%%%%%%%%%%%%%%%%%%%%%%%%%%%%%%%%%%%%%%%%%%%%%%%%%%%%%%%%%%%%%%%%%%

%%% The acknowledgments section is defined using the "acks" environment
%%% (rather than an unnumbered section). The use of this environment 
%%% ensures the proper identification of the section in the article 
%%% metadata as well as the consistent spelling of the heading.

% \begin{acks}
% If you wish to include any acknowledgments in your paper (e.g., to 
% people or funding agencies), please do so using the `\texttt{acks}' 
% environment. Note that the text of your acknowledgments will be omitted
% if you compile your document with the `\texttt{anonymous}' option.
% \end{acks}

%%%%%%%%%%%%%%%%%%%%%%%%%%%%%%%%%%%%%%%%%%%%%%%%%%%%%%%%%%%%%%%%%%%%%%%%

%%% The next two lines define, first, the bibliography style to be 
%%% applied, and, second, the bibliography file to be used.

\bibliographystyle{ACM-Reference-Format} 
\bibliography{sample}

%%%%%%%%%%%%%%%%%%%%%%%%%%%%%%%%%%%%%%%%%%%%%%%%%%%%%%%%%%%%%%%%%%%
%%

\end{document}

%% file: introduction.tex
% Neural networks have outperformed existing approaches in finding an optimal mapping between the given input and output data. The mapping could represent a classification problem \cite{Reinforcement2,ICAByDL}, or regression\cite{Dutting2017,golowich2018deep,manisha2018learning} or even an algorithm\cite{kim2018communicationICLR,graphalogrithmsusingnn}. Given enough data, hyper-parameter tuning, and proper training, the networks are adept at learning effective transformations. For example, in the field of \emph{auction-based resource allocation},  neural networks show impressive performance for finding optimal allocations and payments \cite{cai2018reinforcement,Dutting2017,manisha2018learning,shen2019automated,TacchettiDeepmind,wang2020mechanism}. 
% In this work, we focus on allocating items among interested agents to satisfy certain \emph{fairness} and \emph{efficiency} properties.
% \MP{Remove the first para }

Consider a situation where a social planner needs to allocate a set of indivisible items (goods or/and chores) among interested agents. 
Agents have valuations for the items, i.e., an item might be a \emph{good} -- positive valuation for one while it might be a \emph{chore} -- negative valuation for the other. The agents reveal their valuations upfront to the \emph{social planner.}, The social planner is responsible for the fair and efficient allocation of these items among the agents. 
For example, a Government needs to distribute resources or/and delegate tasks amongst its subdivisions. The subdivisions should not feel mistreated in the system (fairness), and while ensuring fairness, the Government would like to further allocate items optimally for the system's growth (efficiency). 

The fair division of items is well-explored in literature \cite{bouveret_chevaleyre_maudet_moulin_2016, MarkakisTRENDS2017,procaccia_moulin_2016,inbook}. Economists have proposed various fairness notions (Envyfree-ness \cite{Foley1967ResourceAA}, Equitable \cite{EQDubinsSpanier}, Proportional \cite{propSTEIHAUS}) and efficiency notions (Utilitarian Welfare, Nash Welfare, Egalitarian Welfare, Pareto  Efficiency). These are applicable in real-world settings, such as division of investments and inheritance, vaccines, tasks, etc. There are web-based applications such as Spliddit, The Fair Proposals System, Coursematch, Divide Your Rent Fairly, etc., used for credit assignment, land allocation, division of property, course allocation, and even task allotment. All these applications assure certain fairness and efficiency guarantees.

\noindent\textbf{Fairness Notion. }One of the most popular fairness criteria is envy-freeness (EF)\cite{Foley1967ResourceAA}. An allocation is envy-free if each agent values its share at least as much as they value any other agent's share, i.e., no agent is envious. EF is also trivially satisfied by allocating empty bundles to every agent. Hence we must also have efficiency/optimality guarantees. When we consider a complete allocation of indivisible items, Envy-free may not exists—for example, two agents, one good. The agent who doesn't get the good will always envy the one who receives it.
Finding whether EF allocation exists or not is known to be $\Delta^p_2$complete \cite{JAIR3213}, let alone finding an efficient allocation among EF. To overcome this limitation, we consider a prominent relaxation of EF - EF1 (Envy-freeness up to one item) [11].  Unlike EF, EF1 always exists and can be computed in polynomial time \cite{Lipton2004}.

\noindent\textbf{Efficiency Notion. }The notion of Pareto efficiency \footnote{\scriptsize{ An allocation $A$ is said to be PE if it is not Pareto dominated by any other allocation, i.e., there is no other allocation $A'$ such that
$\forall i \in N, v_i(A'_i) \ge v_i(A_i)$, and $\exists i \in N, v_i(A'_i) > v_i(A_i)$.}} is widely studied in fair resource allocation, i.e., PE and fair allocations \cite{barman2018finding,Caragiannis2018,EQ1freemangoods}. In this work, we instead focus on \emph{utilitarian} social welfare (USW), i.e., the sum of utilities of individual agents. When valuations are additive, finding allocations that maximize utilitarian welfare (MUW) is polynomial-time solvable. While finding EF1 or MUW allocations are polynomial-time solvable, maximizing utilitarian welfare within EF1 allocation is an NP-hard problem ~\cite{Aziz2016AAMAS,Caragiannis2018,KeijzerBKZ09,aziz2019constrained,Barman2019Nphard} even in additive valuation settings. 
 There is existing work that provides approximate efficiency and fairness guarantees in \cite{Amanatidis_2017mms, barman2018finding, multiplicity2020arxiv, mmsapprox}. But to find allocations that are MUW among EF allocations is an NP-Hard problem even when valuations are additive for two agents \cite{Barman2019Nphard,ef1inumHaris}. 

Aziz et al.~\cite{ef1inumHaris} provide a pseudopolynomial-time algorithm for finding MUW within EF1, which is exponential in the number of agents and polynomial in the number of items and $V$, where $V$ bounds the valuation per item. When agents valuations are additive and drawn randomly from a uniform distribution, envy-free allocation exists with a high probability when the number of items $m$ is at least $\Omega(n \log n )$ and can be obtained by MUW allocations proven by Dickerson et al.~\cite{dickersonThecomputationalriseandfalloffairness}. This guarantee is achieved only for significantly large values of $m$. However, the hidden constants might be high\footnote{\scriptsize Our experiments show that in the case of uniform distribution, even for 10 agents, 150 items, the probability of MUW allocation being EF1 is less than 0.5}, and it leaves scope to explore. 

With these theoretical limitations, in this paper, we focus on a data-driven approach, i.e., given the agents' valuations, we aim to learn allocations that maximize USW amongst EF1, which we call \emph{EEF1}; efficient and envy-free up to one item.
It is widely known that neural networks outperform existing approaches in finding an optimal mapping (e.g., mechanisms, algorithms) between the given input and output data~\cite{Reinforcement2,ICAByDL,Dutting2017,golowich2018deep,manisha2018learning,kim2018communicationICLR,graphalogrithmsusingnn,cai2018reinforcement,shen2019automated,TacchettiDeepmind,wang2020mechanism}. Given enough data, hyper-parameter tuning, and proper training, the networks are adept at learning effective transformations. 
% For example, in the field of \emph{auction-based resource allocation},  neural networks show impressive performance for finding optimal allocations and payments \cite{cai2018reinforcement,Dutting2017,manisha2018learning,shen2019automated,TacchettiDeepmind,wang2020mechanism}. 
Duetting et al.~\cite{Dutting2017} learn a mechanism from input valuation space to allocations and payments that provide maximum revenue and ensure truthful valuation elicitation.   Motivated by the success of NNs in resource allocations and the theoretical limitations,  we propose a learning-based approach for fair and efficient resource allocations. 
Note that payments are at their disposal in most previous NN-based resource allocation, and the main challenge is to learn payments. In our work, there are no payments, and we are learning allocations via NNs. 
The major challenges are as follows,

\smallskip
\noindent\textbf{Challenges. } To the best of our knowledge, this is the first study that integrates deep learning and fair resource allocation. It entails addressing the following challenges. (i) Allocations for indivisible items are in the discrete space, whereas the output of NNs being real numbers, it can easily learn optimal fractional fair allocations, i.e., give each agent an equal proportion of an item. If we convert fractional solutions to integral solutions in our settings, fairness guarantees no longer hold. 
(ii) Further, we aim to design a generalized network that should work for any number of agents or items, even for configurations not seen during training. Most of the existing NN based approaches in EconCS, train the models separately for each configuration, for example, in papers \cite{manisha2018learning,Dutting2017} 

\smallskip
\noindent \textbf{Contributions.}
For a given distribution, there is a certain likelihood for MUW allocation to be EF1. It increases when the number of items is very large, w.r.t. the number of agents. We train the NN for the cases where MUW allocation is rarely EF1. In order to provide competitive results, we address the above challenges by introducing the following novelty,
\begin{itemize}
    \item  We convert the valuations into a six-channeled input to the network for better performance.
    \item  We have a series of convolutional and up-convolutional layers to learn EEF1 allocations. Since the network is fully convolutional, it is generalized for any number of agents and items once trained accordingly.
    \item  We formulate a Lagrangian loss function to find allocations that are \emph{Envy Free up to one item Efficient (maximal USW)} and EF1.
    \item  We show that, for our setting, the Bagging of multiple networks improves performance.
    \item  We sample valuations from various distributions and report the expected fairness and efficiency achieved. Our network performs well even for large instances with more than 10 agents and 100 items. Moreover, compared to any integer optimization solver, the network quickly computes the output; hence we can improvise this approach to be adept in practical real-time applications.
\end{itemize}